\documentclass[
 floatfix,
 reprint,
 amsmath,amssymb,
 aps,
]{revtex4-2}
\usepackage[english]{babel}
\usepackage[utf8]{inputenc}
\usepackage[T1]{fontenc}
\usepackage{physics}
\usepackage[english]{babel}
\usepackage{xspace,amsmath}
\usepackage{amssymb,braket,bbm}
\usepackage{hepparticles,graphicx}
\usepackage[top=2cm,left=2cm,right=2cm]{geometry}
\usepackage{color}
\usepackage{mathrsfs}
\usepackage{comment}
\usepackage{physics}
\usepackage{xcolor}
\usepackage{amsmath}
\usepackage{graphicx}
\usepackage{svg}
\usepackage{dcolumn}
\usepackage[colorlinks=true,allcolors=blue]{hyperref}

\newcommand{\pardd}[2]{\frac{\partial{#1}}{\partial{#2}}}
\newcommand{\kbar}{\bar{\kappa}}

\begin{document}
\title{A bound on the free energy of tensionless membranes}

\author{Francesco Serafin}
\affiliation{Department of Physics, University of Michigan, Ann Arbor, MI 48109-1040, USA}%
\author{Mark J. Bowick} 
\affiliation{Kavli Institute for Theoretical Physics, University of California Santa Barbara}

\begin{abstract}
    Using the proof of Willmore's conjecture \cite{Willmore65} by Marques and Neves \cite{Marques2014}, we conjecture that the free energy of tensionless fluid membranes of arbitrary genus has an upper bound. This implies that the average genus of such a membrane, in equilibrium, is finite, regardless of external constraints. We propose that the Gaussian rigidity may be determined by measuring the relative frequencies of large-genus configurations at low temperature.
\end{abstract}

\maketitle

\section{Coarse-grained energy of the membrane}
Fluid membranes are found in a wide variety of physical settings. Spherical shapes (genus 0) and higher genus surfaces (see e.g. \cite{MutzBensimon91} for genus 1, \cite{FMB_PhysRevLett.68.2551} for genus 2 and \cite{high-genus-expt,high-g-expt-2,Dogic,Michalet94} for higher genus) have been observed in experiments with unpolymerized phospholipid membranes, as well as in numerical simulations \cite{Julicher_morpho}.
\par We treat here the ensemble of all membrane shapes at a scale $L\gg a$ much larger than the molecular size $a$, justifying the use of a standard continuum model.  The membrane shape will be modeled as a 
2-dimensional surfaces of genus $g$ embedded in $\mathbb{R}^3$ $\mathbf{X}_g(\boldsymbol{\xi})$, with internal coordinates $\boldsymbol{\xi}=(\xi^1,\xi^2)$. The subscript $g$ labels the number of handles (genus) of the surface $\Sigma$. The total energy associated to a configuration is 
\begin{equation}
  \label{eq:totalHamiltonian}
    \mathcal{H}[\mathbf{X}_g]=\int_{\Sigma_g} \left[ \gamma+ \frac{\kappa}{2}  \vec{H}^2 + \bar{\kappa}  K_\mathrm{G} \right]\, \mathrm{d}S \quad,
\end{equation}
where $\mathrm{d}S=\sqrt{\det g_{ab}}\, \mathrm{d}^2\xi $ and $g_{ab}(\boldsymbol{\xi})= \partial_a\mathbf{X}(\boldsymbol{\xi})\cdot \partial_b\mathbf{X}(\boldsymbol{\xi})$ is the induced metric tensor.  $\vec{H}$ and $K_\mathrm{G}$ are the mean and Gaussian curvatures, respectively:
\begin{align}
\vec{H}^2(\boldsymbol{\xi})&=\frac{1}{4}\left(\frac{1}{R_1(\boldsymbol{\xi})}+\frac{1}{R_2(\boldsymbol{\xi})}\right)^2 \quad ,\\ K_\mathrm{G}(\boldsymbol{\xi})&=\frac{1}{R_1(\boldsymbol{\xi})R_2(\boldsymbol{\xi})} \quad.
\end{align}
Here $R_{1,2}(\boldsymbol{\xi})$ are the local principal curvature radii of the surface \cite{docarmo}. We restrict ourselves to the symmetric case for which the spontaneous curvature vanishes.

\par The first term in \eqref{eq:totalHamiltonian} is an area term controlled by the surface tension $\gamma$. For a membrane coupled to a reservoir of constituents the surface tension will vanish in equilibrium to minimize the energy \cite{Taupin,Leibler}. In the following we will work in this ensemble and thus set $\gamma=0$. The second term in \eqref{eq:totalHamiltonian} is the bending energy, measuring the cost of shape change, and is controlled by a bending rigidity $\kappa$, with dimensions of energy. 
\begin{equation}
\label{eq:Willmore}
    \mathcal{W}[\mathbf{X}_g]\equiv\int_{\Sigma_g}    \vec{H}^2\, \mathrm{d}S
\end{equation}
is known as the Willmore functional. It was first proposed by Poisson and Germain \cite{germain_2013} as an energy for elastic shells.  The stationary points of $\mathcal{W}$ are called Willmore surfaces. Remarkably, the Willmore functional is invariant under M\"obius transformations of $\mathbb{R}^3$ \cite{WillmoreProceedings,LiYau,DUPLANTIER1990179} and so the degenerate minima of $\mathcal{W}$ are related by conformal transformations of $\mathbb{R}^3$. 

The term $\bar{\kappa}K_\mathrm{G}$ in \eqref{eq:totalHamiltonian} is the energy cost of Gaussian curvature fluctuations, where $\bar{\kappa}$ is known as the Gaussian rigidity.  By Gauss-Bonnet it is a topological invariant for $\Sigma_g$ a closed surface
\begin{equation}
    \int_{\Sigma_g}    K_\mathrm{G}\, \mathrm{d}S=2\pi\chi({\Sigma_g}) \quad, \quad \chi=2(1-g) \,.
\end{equation}
The last two terms of \eqref{eq:totalHamiltonian} together are usually known as the Helfrich-Canham energy of fluid membranes \cite{Helfrich,CANHAM197061}:
\begin{equation}
\label{eq:HCenergy}
    \mathcal{H}_\mathrm{HC}\equiv \frac{\kappa}{2}\mathcal{W}[\mathbf{X}_g]+4\pi\bar{\kappa}(1-g) \quad.
\end{equation}
For $\bar{\kappa}>0$ it is energetically favorable to form an arbitrarily large number of handles ($g\to\infty$) {and \eqref{eq:HCenergy} would be unbounded below}. In a physical system steric {effects} and other constraints will, however, limit the formation of very high genus surfaces~\footnote{See also \cite{Gaussian_curv_mono} for the sign of $\bar{\kappa}.$}. {In essence the effective $\bar{\kappa}$ is likely to be negative.}
\section{Statistical ensemble of tensionless fluid membranes}
\par We consider the statistical ensemble of all possible configurations of the surface $\mathbf{X}_g$ at equilibrium at temperature $T$ and zero surface tension \cite{Leibler}. The canonical partition function $Z$ is 
\begin{equation}
\label{eq:Z}
    Z=\int_{\mathrm{all \,configs.}} \mathscr{D}\mathbf{X}\, e^{-\beta \mathcal{H}_\mathrm{HC}[\mathbf{X}]} \quad,
\end{equation}
where the sum is over connected as well as disconnected surfaces of arbitrary topology and $\beta=1/(k_\mathrm{B}T)$ with $k_\mathrm{B}$ the Boltzmann constant. We will work with the Helmoltz free energy $F=-k_\mathrm{B}T\log Z$, by restricting the sum in \eqref{eq:Z} to \textit{connected} configurations (two disjoint spheres, e.g., would not be counted).
%The number of components $N$ used in \cite{Julicher_morpho} is fixed
Then the sum over closed, compact, orientable embedded surfaces can be organized into an integration $\mathscr{D}\mathbf{X}_g$ over fluctuations at fixed genus $g$ and a sum over all genera
\begin{equation}
\label{eq:freeenergy}
    F=\sum_{g=0}^\infty \int_\mathrm{conn.} \mathscr{D}\mathbf{X}_g e^{-\beta \mathcal{H}_\mathrm{HC}[\mathbf{X}_g]} \quad.
\end{equation}
\begin{figure}
    \centering
    \includegraphics[width=0.4\textwidth]{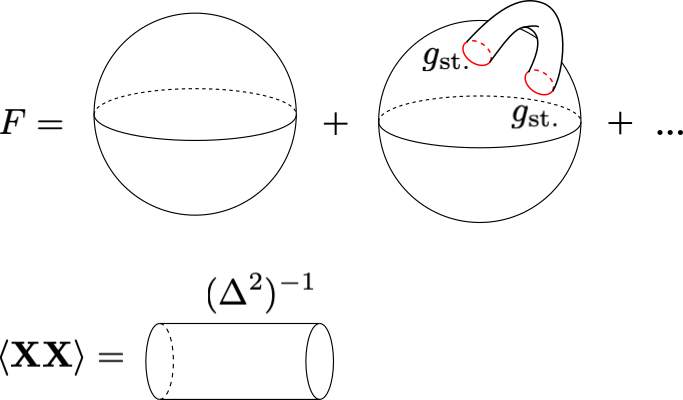}
    \caption{\emph{Top.} Diagrammatic representation of the free energy~\eqref{eq:F1} of tensionless fluid membranes. The first terms in the sum are topological spheres ($g=0$), tori with 1 handle ($g=1$) and so on. The genus increases by one unit when a handle is attached. Every handle contributes with two factors of the coupling constant $g_\mathrm{st.}$ (defined in the main text), one for each excised disk (in red). \emph{Bottom.} The free propagator (the inverse biharmonic operator) is represented as an open cylinder.}
    \label{fig:Feyn}
\end{figure}

The first term in the sum counts all topological spheres ($g=0$), the second all tori with one handle ($g=1$), and so on. Using \eqref{eq:HCenergy} in \eqref{eq:freeenergy} and factorizing those terms that do not depend on the geometry, we can write
\begin{equation}
\label{eq:F1}
    F=e^{-4\pi\beta\bar{\kappa}}\sum_{g=0}^\infty e^{4\pi\beta\bar{\kappa}g}\int_\mathrm{conn.} \mathscr{D}\mathbf{X}_g \,e^{- \frac{\beta\kappa}{2}\mathcal{W}[\mathbf{X}_g]} \quad.
\end{equation}
We interpret $ \mathcal{H}_\mathrm{HC}[\mathbf{X}_g]$ as the Euclidean action of the field $\mathbf{X}$ and $F$ as the generating function of connected diagrams for the $n-$point correlators $\braket{X^\mu(\boldsymbol{\xi}_1)...X^\nu(\boldsymbol{\xi}_n)}$ where $\boldsymbol{\xi}$ are the internal coordinates of the surface $\mathbf{X}$. Then each configuration contained in the expansion~\eqref{eq:F1} is analogous to a Feynman diagram of a closed-string amplitude whose worldsheet is a closed surface of genus $g$ weighted by the factor $(\exp{2\pi\beta \bar{\kappa}})^{2g}$ (see Fig.~\ref{fig:Feyn},\emph{Top}).

Starting from a configuration of genus $g$, the subsequent term in the sum is obtained by attaching a handle. Each handle contributes with a factor $g_\mathrm{st}^2$, where $g_\mathrm{st}\equiv(\exp{2\pi\beta \bar{\kappa}})$ is analogous to the string coupling constant in closed string theory. The power of 2 appears because every handle is a cylinder anchored to the surface at \emph{two} excised disks (see Fig.~\ref{fig:Feyn},\emph{Top}). The probability of forming a handle is controlled by both the temperature $\beta$ and the Gaussian rigidity $\Bar{\kappa}$. When $\Bar{\kappa}<0$ the formation of handles is suppressed ($g_\mathrm{st}<1$).

In the partition function~\eqref{eq:Z} the role of the Euclidean action is played by the Willmore energy~\eqref{eq:Willmore}. The latter can be written as $\mathcal{W}[\Sigma_g]=\int (\Delta \mathbf{X})^2\,\mathrm{d}S$ and is a fourth-order operator in $\partial_a$. Thus, the free propagator $\braket{X^\mu(\boldsymbol{\xi}_1)X^\nu(\boldsymbol{\xi}_2)}$ (represented by a cylinder in Fig.~\ref{fig:Feyn},\emph{Bottom}) is the inverse of the \emph{biharmonic} operator: $(\Delta\Delta)^{-1}$, where $\Delta=\partial_a\partial^a$ is the Laplacian. This is different from the propagator of closed strings: there, the Polyakov action is quadratic in the derivatives ($S_\mathrm{P}=\int (\partial_a\mathbf{X})^2\,\mathrm{d}S$) so the free propagator is the inverse of the Laplacian $\Delta^{-1}$.

\section{Upper bound conjecture}
We ask: is $F$ [Eq.~\eqref{eq:F1}] finite or divergent? The answer depends on the asymptotics of the functional integral as a function of $g$. For topological spheres $S^2$ ($g=0$) the Willmore functional is bounded below by $4\pi$:
\begin{equation}
\label{eq:ws}
    \mathcal{W}[S^2]\geq 4\pi \quad .
\end{equation}
The bound is saturated for the round sphere of radius $R$, $S^2(R)$ modulo uniform scaling of the sphere radius. A simple rescaling does not change shape, so we regard all round spheres as the same configuration. With this caveat, the first term in the series is bounded above by $e^{-4\pi\beta\bar{\kappa}}e^{-(\beta\kappa/2)4\pi}$. 

For $g\geq 1$ in \eqref{eq:F1} we must consider all topological tori.
In 1965 T. Willmore \cite{Willmore65} conjectured that $\mathcal{W}\geq 2\pi^2$ for all genus-1 surfaces and that the Clifford torus~\footnote{The embedded Clifford torus is the stereographic projection in $\mathbb{R}^3$ of the unit 3-sphere's equator $(\sqrt{2})^{-1}\{\cos\varphi,\sin\varphi,\cos\theta,\sin\theta\}$.} saturates the bound up to conformal transformations of $S^3$. In 2014 Marques and Neves \cite{Marques2014} proved that for all \emph{embedded} closed surfaces of genus $g\geq1$ 
\begin{equation}
\label{eq:wc}
\mathcal{W}[\mathbf{X}_{g\geq1}]\geq 2\pi^2   \quad, 
\end{equation}
with equality holding only for the Clifford torus (up to conformal transformations), thus proving Willmore's conjecture. 
{At the time of writing, there is ongoing research   \cite{KusnerComparison,Lawson_minima} aiming to prove that Lawson's minimal surfaces (denoted $\xi_{g,1}$ \cite{lawson_surface_ann_math}) of genus $g\geq2$ are the minimizers of the Willmore functional among all surfaces with the same ambient symmetries as $\xi_{g,1}$.}
%The generalized Willmore conjecture for surfaces without special symmetries and $g\geq2$ remains open \mjb{the generalized Willmore conjecture has not been defined and I don't think we use this result in any way so it seems gratuitous}. 

Using \eqref{eq:ws} and \eqref{eq:wc} in \eqref{eq:F1} 
\begin{equation}
\label{eq:F2}
    F\leq e^{-4\pi\beta\bar{\kappa}}\left\{e^{-2\pi\beta\kappa} +e^{- 2\pi^2\beta\kappa} \sum_{g=1}^\infty e^{4\pi\beta\bar{\kappa}g}\int \mathscr{D}\mathbf{X}_g \right\}
\end{equation}
The integral over geometric fluctuations $\Gamma(g)\equiv\int \mathscr{D}\mathbf{X}_g $ at fixed genus is subtle to evaluate because it requires a parametrization of the moduli space of \emph{embedded} surfaces. Moduli spaces of non-embedded surfaces lead to divergent partition functions - as happens in string theory \cite{Gross}. Here we are concerned with the moduli space of physical, \emph{embedded} membranes. This constraint may lead to a finite growth of moduli space and a convergent free energy. From a physical point of view, as noted earlier, there will be natural suppression of higher-genus contributions arising both from steric constraints (handles have a minimal physical size) and the full amplitude for creating a handle, here parametized by a single coupling constant $\bar{\kappa}$. 

In a low temperature expansion one sums over slight deformations of the ground states of the Willmore functional. For each genus there is a continuous family of ground states related by conformal transformations, thanks to the conformal invariance of \eqref{eq:Willmore}. Since the conformal group contains unbounded elements it is necessary to sum over equivalence classes of shapes modulo rotations, translations and uniform dilatation which preserve the shape, as noted eaerlier for the sphere ($g=0$). One must then determine if the integration over special conformal transformations (SCTs) is finite.

As an example, we consider the embedded axisymmetric Clifford torus
\begin{equation}
\label{eq:cliff}
    \mathbf{X}_\mathrm{Cl}(\theta,\varphi)=\frac{1}{\sqrt{2}-\sin\varphi}\left(\cos\theta,\sin\theta,\cos\varphi\right) 
\end{equation}
where $\theta,\varphi\in[0,2\pi[$
and we apply a SCT
\begin{equation}
\label{eq:SCT}
 \mathbf{X}\to \frac{\mathbf{X}/X^2+\mathbf{a}}{(\mathbf{X}/X^2+\mathbf{a})^2}    \quad.
\end{equation}
\begin{figure}
    \centering
    \includegraphics[width=8cm]{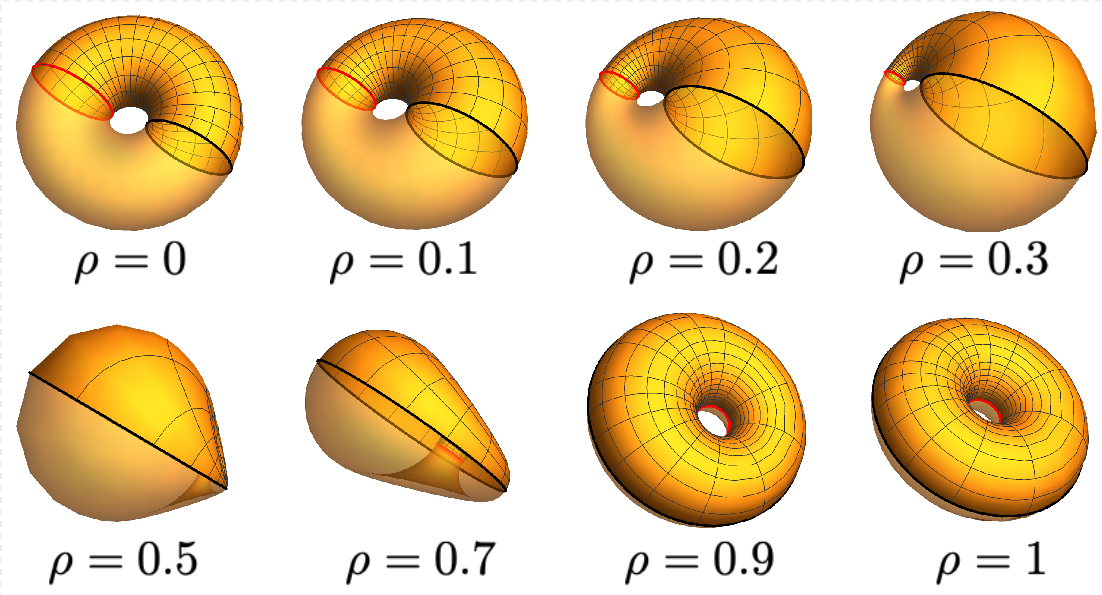}
    \caption{Family of ground states of $\mathcal{W}$ among genus 1 surfaces, generated by applying \eqref{eq:SCT} with $\mathbf{a}=(\rho,0,0)$ to the axisymmetric Clifford torus \eqref{eq:cliff} at $\rho=0$. Upon increasing $\rho$ \eqref{eq:SCT} the circle marked in red shrinks while the black circle inflates. The black circle's radius diverges as $\rho\to\sqrt{2}-1\simeq0.414$. The shapes repeat for $\rho>\sqrt{2}-1$ up to a global scaling.}
    \label{fig:fig0}
\end{figure}
The Willmore energy of the transformed shape is still equal to $2\pi^2$. Moreover, it is known \cite{Julicher,FourcadeJPhysII} that $\mathbf{a}=(0,0,a_z)$ leads to a simple rescaling of \eqref{eq:cliff} while  $\mathbf{a}=\rho(\cos\alpha,\sin\alpha,0)$ generates a one-parameter family of shapes (for $0<\rho<\sqrt{2}-1$) that break axial symmetry. $\alpha$ simply rotates the shape around the $z-$axis. This space $\Gamma(1)$ is \textit{bounded} \cite{Julicher} because at $\rho=\sqrt{2}-1$ the axisymmetric shape \eqref{eq:cliff} deforms into a sphere with an infinitesimal handle. If $\rho$ increases further the transformation retraces the shapes that appeared between $0<\rho<\sqrt{2}-1$, so at least for $g=1$ the degeneracy factor $\Gamma(1)$ represents a bounded family of genus 1 shapes.
We conjecture that similar arguments hold for higher-genus surfaces. If $\Gamma(g)$ is finite for all $g$, it is crucial to know how $\Gamma(g)$ scales with $g$. We comment on some representative cases below.

\par If $\Gamma=\Gamma_0$ is independent of $g$, the infinite sum in \eqref{eq:F2} can be rewritten as a geometric series. If $e^{4\pi\beta\bar{\kappa}}<1$ \footnote{The condition for convergence means that the free energy~\eqref{eq:F1} is a perturbative expansion in $g_\mathrm{st.}$} it is possible to sum the series in \eqref{eq:F1} and the bound takes the form
\begin{equation}
\label{eq:bound1}
    F\leq e^{-4\pi\beta\bar{\kappa}}e^{-2\pi\beta\kappa} + \frac{\Gamma_0 e^{- 2\pi^2\beta\kappa}}{1-e^{4\pi\beta\bar{\kappa}}} \quad.
\end{equation}
The result \eqref{eq:bound1}, with $\Gamma_0=1$, becomes exact for an ensemble of closed membranes with fixed isoperimetric ratio, defined as $r=V^2/A^3$ where $V,A$ are the volume and area of the closed surface. When $r$ is fixed, there is a unique ground state for the Willmore functional.
\par If $\Gamma(g)$ is a polynomial, say $\Gamma(g)= A g^k$, then the bound reads
\begin{equation}
    F\leq e^{-4\pi\beta\bar{\kappa}}\left\{e^{-2\pi\beta\kappa} +A e^{- 2\pi^2\beta\kappa} \mathrm{Li}_{-k}(e^{4\pi\beta\bar{\kappa}})\right\}
\end{equation}
where $\mathrm{Li}_k(z)$ is the polylogarithmic function \cite{Lewin:1981:PAF}.
\section{Bound on the average number of handles}
Consider a connected membrane with many handles such as a microemulsion in the plumber's nightmare phase \cite{Huse-Leibler}. The conjectured bound \eqref{eq:F2} has observable implications for the expected number of handles. Let $\braket{g}$ be the expectation value of the genus
\begin{equation}
    \braket{g}\equiv\frac{1}{Z}\sum_{g=0}^\infty  g \cdot e^{-4\pi\beta \bar{\kappa}(1-g)} \int \mathscr{D}\mathbf{X}_g  \,e^{- \frac{\beta\kappa}{2}\mathcal{W}[\mathbf{X}_g]} \quad.
\end{equation}
This can be written in terms of the free energy $F=-\beta \log Z$ as
\begin{equation}
\label{eq:genusbound1}
    \braket{g(\kbar)}=1-\frac{1}{4\pi}\pardd{F(\kbar)}{\bar{\kappa}} \quad,
\end{equation}
where the free energy and its conjectured upper bound depend parametrically on the Gaussian rigidity $\bar{\kappa}$. Suppose that the value of $\kbar$ could be varied in experiments, for instance by changing the nature of the surfactants - see for example \cite{Deserno2} and Table 1 in \cite{Deserno}) and that the associated value of $\braket{g(\kbar)}$ could be measured for each realization of the ensemble \cite{measur_KG}.
Using \eqref{eq:genusbound1}, the  ensemble average of $\braket{g(\kbar)}$ is 
\begin{align}
\label{eq:genusbound2}
\nonumber
   \overline{\braket{g(\kbar)}}&\equiv \frac{1}{\Delta\kbar} \int_{\kbar_1}^{\kbar_2} \, \braket{g(\kbar')} \,\mathrm{d}\kbar' \\ &= 1-\frac{1}{4\pi\Delta\bar{\kappa}}[F({\kbar_2})-F({\kbar_1})]
\end{align}
where we assumed that the Gaussian rigidity can be varied in the range $\kbar_1<\kbar<\kbar_2$ and $\Delta \kbar\equiv\kbar_2-\kbar_1$.
Let $b(\bar{\kappa})$ indicate the conjectured upper bound \eqref{eq:F2} on $F$: $F(\kbar)\leq b(\kbar)$.  Then equation \eqref{eq:genusbound2} turns into the following bound:
\begin{equation}
\label{eq:genusbound3}
    \overline{\braket{g(\kbar)}}<1-\frac{1}{4\pi\Delta\bar{\kappa}}[b(\kbar_2)-b(\kbar_1)] \quad.
\end{equation}
The average on the left-hand side in \eqref{eq:genusbound3} could be measured in experiments. The bound \eqref{eq:genusbound3} predicts that $\overline{\braket{g(\kbar)}}$ is bounded above and so the expected number of handles should be finite and should depend only on the interval $\Delta\kbar$. Since $g\geq0$, \eqref{eq:genusbound3} implies that $b(\kbar)$ is a non-increasing function of the Gaussian rigidity. Eq.~\eqref{eq:genusbound3} implies that a tensionless fluid membrane at equilibrium has a finite average genus regardless of other interactions or constraints such as self-avoidance and steric repulsion.

\section{Probability of forming a handle}
Using Eq.~\eqref{eq:Z} and \eqref{eq:HCenergy}, the probability that a membrane of genus $g$ occurs at equilibrium is 
\begin{equation}
    p_g=\sum_{\{\Sigma_g\}}\frac{1}{Z}{e^{-\beta\frac{\kappa}{2} \mathcal{W}[\Sigma_g]-\beta\Bar{\kappa}4\pi(1-g)}} \quad,
\end{equation}
where the sum $\sum_{\{\Sigma_g\}}$ is over all membrane configurations at fixed genus. We compute the ratio between $p_{g+1}$ and $p_g$. Since $Z$ and $\exp{-
\beta\Bar{\kappa}4\pi(1-g)}$ don't depend on the geometry of the configuration we find
\begin{equation}
    \frac{p_{g+1}}{p_g}=e^{\beta 4 \pi \Bar{\kappa}}\frac{\sum_{\{\Sigma_{g+1}\}}e^{-\beta\frac{\kappa}{2} \mathcal{W}[\Sigma_{g+1}]}}{\sum_{\{\Sigma_{g}\}}e^{-\beta\frac{\kappa}{2} \mathcal{W}[\Sigma_{g}]}} \quad.
\end{equation}
At low temperature ($\beta\to\infty$) we expect that the sum will be dominated by those configurations that minimize the Willmore functional:
\begin{equation}
\label{eq:ratiog}
      \lim_{\beta\to\infty}  \frac{p_{g+1}}{p_g}\sim e^{\beta 4 \pi \Bar{\kappa}}\frac{e^{-\beta\frac{\kappa}{2} \inf\mathcal{W}[\Sigma_{g+1}]}}{e^{-\beta\frac{\kappa}{2} \inf\mathcal{W}[\Sigma_{g}]}} \quad,
\end{equation}
where $\inf$ denotes the infimum. First, we restrict our attention to membranes of genus 0 (i.e. spheres, for which  $\inf\mathcal{W}[\Sigma_0]=4\pi$) and 1 (i.e. tori, for which $\inf\mathcal{W}[\Sigma_1]=2\pi^2$). Then Eq.~\eqref{eq:ratiog} becomes
\begin{equation}
\label{eq:ratio10}
      \lim_{\beta\to\infty}  \frac{p_{1}}{p_0}\sim e^{\beta 4 \pi \Bar{\kappa}}e^{-\beta\kappa(\pi^2-2\pi)} \quad.
\end{equation}
The left-hand side of Eq.~\eqref{eq:ratio10} could be determined experimentally by measuring the frequency of topological spheres and genus-1 tori at equilibrium and low temperature. If the bending rigidity $\kappa$ could be measured independently, the Gaussian rigidity $\Bar{\kappa}$ could be determined from Eq.~\eqref{eq:ratio10}.
\par Another relation between relative probabilities and the Gaussian rigidity may be found by using R. Kusner's result~\cite{KusnerComparison} that the infimum of the Willmore's functional is bounded above by $8\pi$ for very large genus: $\lim_{g\to\infty}\inf\mathcal{W}[\Sigma_g]=8\pi$. In the large-genus limit both numerator and denominator in the right-hand side of Eq.~\eqref{eq:ratiog} tend to the same value $\exp(-\beta\kappa/2\cdot8\pi)$ and the ratio $p_{g+1}/p_g$ becomes independent of $\kappa$:
\begin{equation}
\label{eq:ratioginfty}
    \lim_{\beta\to\infty}\lim_{g\to\infty}  \frac{p_{g+1}}{p_g}\sim e^{\beta 4 \pi \Bar{\kappa}} \quad.
\end{equation}
Equation~\eqref{eq:ratioginfty} suggests that $\Bar{\kappa}$ could be determined directly by measuring the relative frequency of large-genus configurations (which may be rarer than spheres and genus-1 tori) at low temperature. This could provide a new method for finding $\Bar{\kappa}$, whose value is notoriously difficult to determine~\cite{gaussianmodulus,Gaussian_curv_mono}.
\section{Conclusions}
We considered the free energy of closed, single-component membranes with vanishing surface tension. Using Willmore's bound on the bending energy we conjectured that the free energy is bounded above by a universal constant that depends on the bending rigidity and on the Gaussian rigidity. Such a bound implies that the membrane has a finite average number of handles. 
{By interpreting the partition function as a field theory of fluid membranes we observed that the formation of handles is controlled by an effective coupling constant that depends on the Gaussian rigidity and the temperature. Finally, using the well-known upper bound on the infimum of the Willmore energy~\cite{KusnerComparison}, we propose that the Gaussian rigidity $\Bar{\kappa}$ may be determined in experiments by measuring the relative frequencies of large-genus configurations at low temperature.}
\section{Acknowledgements}
This research was supported in part by the National Science Foundation under Grant No. NSF PHY-1748958. We benefited from several stimulating discussions with Rob Kusner.

\bibliographystyle{amsplain}
\bibliography{Bibliography}
\end{document}